\documentstyle[preprint,aps]{revtex}
\begin{document}
\draft
\title{
Renormaliztion-Group Approach to Spin-Wave Theory of Quantum
Heisenberg Ferromagnet
	 }
\author{Naoum Karchev\cite{byline}}
\address{
Department of Physics, University of Sofia, 1126 Sofia, Bulgaria
	  }
\maketitle
\begin{abstract}
The renormalization-group method is used to analyze the low-temperature
behaviour of a two-dimentional, spin-$s$ quantum Heisenberg ferromagnet. A set
of recursion equations is derived in an one-loop approximation. The
low-temperature asymptotics of the correlation length and the uniform
susceptibility are obtained. For small spins ($s= 1/2,1$) the results are
essentialy different from those in the spin-wave theory.
\end{abstract}

The revival of the interest in the theory of quantum magnetism has led to
creation of new approaches of investigation. An effective continuum field
theory, which is quantum-mechanical generalization of the classical
nonlinear $\sigma$-model was derived from lattice, large spin Heisenberg
model of antiferromagnets \cite{1}. The two dimensional antiferromagnet on a
square lattice was treated also by means of the Schwinger boson representation
of the spin algebra. This representation allows an appropriate mean-field
theory of the low-temperature behaviour of the sytem \cite{2}. At the same time,
a modified spin-wave theory of Heisenberg (anti)ferromagnets was formulated
\cite{3,4}. The usual spin-wave theory was suplemented with the
constraint that the
magnetization at each site is zero. This ensures that the sublattice
rotational symmetry is not broken.

The results obtained in the papers are in quantitative agreement, but, as
was recognized by the authors, it is difficult to continue them to smaller
values of the spin.

The aim of the present investigation is to obtain a better understanding
of a small spin quantum ferromagnet. The renormalization-group method
is used to analyze the low-temperature behaviour of a two dimensional,
spin-$s$ quantum Heisenberg ferromagnet. A set of recursion equations
is derived in an one-loop approximation. The spin-recursion relation shows
that the effective spin increases (the spin rescaling factor is two) and
hence in the limit of infinitely many recursions, the spin-wave
approximation can be used. The low-temperature asymptotics of the
correlation length and the uniform susceptibility are obtained solving
the recursion relations. For small spins ($s= 1/2,1$) the results are
essentialy different from those in the spin-wave theory.

The spin-$s$ quantum Heisenberg ferromagnet is defined by the hamiltonian
\FL
\begin{equation}
\hat h\,=\,-J\sum\limits_{<i,j>} \hat {\vec S_i}\cdot\hat {\vec S_j}
\end{equation}
where $\hat {\vec S_i}$ are spin operators on site $i(j)$ of a two
dimentional
square lattice with number of sites $N$ and lattice spacing $a$. By $<i,j>$
I denote the sum over the nearest neighbours.

To preserve the sublattice rotational symmetry in spin-wave theory of the
disordered phase, Takahashy \cite{3} imposed a condition of zero
sublattice magnetization $\langle \hat S_i^3 \rangle \,=\,0$.
Using Holstein-Primakoff representation for the spin operators, one can
rewrite it as a condition that the total number, on average, of
spin-waves per site is $s$. To enfors the constraint, one introduces a
new term in the hamiltonian $\hat h\to\hat h\,-\,\mu\sum\limits_i\hat S_i^3$,
with the chemical potential $\mu$ to be determined from the condition.
The system is disordered at any temperature $T\neq 0$ and the chemical
potential is positive $\mu(T)>0$. When temperature goes to zero $\mu(T)$
reaches $0$ and the spin-wave bosons Bose-condense at zero wave-vector.
Due to Bose
condensation at $T=0$, the long-range ferromagnetic correlation is present.

The low-temperature features are determined by the behaviour of the system
near the origin of the Brillouin zone. Following the renormalization group
method one devides the Brillouin zone into two parts. Then the half of the
zone which contains small wave-vectors is of interest to us, and the theory
should be reformulated using the reduced Brilloun zone. In the site
representation, this is equivalent to introduce two sublattices, $A$ and $B$.
I identify the sublattices and refer to the new lattice, with number
of sites\, $N_1=\frac 12 N$\, and lattice spacing\, $a_1=\sqrt {2}a$\,,
as $L_1$. The hamiltonian (1) can be rewritten in the form
\FL
\begin{equation}
\hat h\,=\,-J\sum\limits_{\nu}\sum\limits_{i\in L_1}\left(
\hat {\vec S_i}^A\cdot\hat {\vec S}_{i+a_{\nu}}^B\,+\,
\hat {\vec S_i}^B\cdot\hat {\vec S}_{i+a_{\nu}}^A\right)\,-\,
\mu\sum\limits_{i\in L_1}\left(\hat {S_i}^{A3}\,+\,\hat {S_i}^{B3}\right)
\end{equation}
where the sum is over the two directions $a_x=(a,0)$ and $a_y=(0,a)$ and
over the sites of the new lattice $L_1$. The spin operators $\hat {\vec S_i}^A$
and $\hat {\vec S_i}^B$ can be realized using canonical bose creation and
annihilation operators $\hat a_i^{A+}, \hat a_i^A$ and
$\hat a_i^{B+}, \hat a_i^B$ (Holstein-Promakoff representation)
Then, to leading order of $s^{-1}$ the hamiltonian (2) takes the form
\FL
\begin{eqnarray}
\hat h & = & sJ\sum\limits_{\nu}\sum\limits_{i\in L_1}
\left (\hat a_i^{A+}\hat a_i^A+\hat a_{i+a_{\nu}}^{A+}\hat a_{i+a_{\nu}}^A+
\hat a_i^{B+}\hat a_i^B+\hat a_{i+a_{\nu}}^{B+}\hat
a_{i+a_{\nu}}^B -\hat a_{i+a_{\nu}}^{B+}\hat a_i^A-\hat a_i^{B+}\hat
a_{i+a_{\nu}}^A\right. \\
& & \left.
-\hat a_{i+a_{\nu}}^{A+}\hat a_i^B-\hat a_i^{A+}\hat a_{i+a_{\nu}}^B\right)
+\mu\sum\limits_{i\in L_1}\left (\hat a_i^{A+}\hat a_i^A+
\hat a_i^{B+}\hat a_i^B-2s\right)-2Js^2N \nonumber
\end{eqnarray}

The hamiltonian (3) is diagonalized by the canonical transformation
\FL
\begin{eqnarray}
\hat a_i^A = \frac {1}{\sqrt 2}\left (\hat\alpha_i+\hat\beta_i\right),
\hspace{0.1cm}
\hat a_i^{A+} = \frac {1}{\sqrt 2} \left (\hat\alpha_i^++\hat\beta_i^+\right),
\hspace{0.1cm}
\hat a_i^B = \frac {1}{\sqrt 2}\left (\hat\alpha_i-\hat\beta_i\right),
\hspace{0.1cm}
\hat a_i^{B+} = \frac {1}{\sqrt 2}
\left (\hat\alpha_i^+-\hat\beta_i^+\right),
\end{eqnarray}
In momentum space the wave-vectors run the half of the Brillouin zone.
The dispersion $\varepsilon_k^{\beta}$ of the $\hat \beta_i^+, \hat \beta_i$
modes is positive even at $T=0$ ($\mu=0$) and I
refer to them as "fast modes". On the other hand,
at zero temperature, $\varepsilon^{\alpha}_{k=0}=0$, and I refer to
$\hat\alpha_i^+$ and $\hat\alpha_i$ as "slow modes".

To proceed further it is more convenient to represent the partition function
in terms of path integral. In Matsubara representation
\FL
\begin{equation}
{\cal Z} = \int d\mu (a^{A+}, a^A) d\mu (a^{B+}, a^B)
\exp \left[-\int\limits_0^{\beta}d\tau\left[\sum\limits_{i\in L_1}
\left (a_i^{A+}\dot a_i^A+a_i^{B+}\dot a_i^B\right)+h(\tau)\right]\right]
\end{equation}
In formula (5) $\beta$ is the inverse temperature,
$a_i^{A+}(\tau),a_i^A(\tau)$ and $a_i^{B+}(\tau),a_i^B(\tau)$
are complex functions on imaginery time $\tau$, subject to conditions
$a_i^{A+}(\tau)a_i^A(\tau)\leq 2s$ and $a_i^{B+}(\tau)a_i^B(\tau)\leq 2s$, and
$d\mu(a^{R+},a^R)$ is the usual path integral measure for bosons. The fields
obey periodic boundary conditions. The hamiltonian
$h(\tau)$ is obtained from Eqs (2) replacing the creation and
annihilation operators by the complex fields.
In particular, the spin operators are replaced by the vectors
$\vec S_i^A(\tau)$ and $\vec S_i^B(\tau)$ that satisfy $\vec S_i^A(\tau)^2=s^2$
and $\vec S_i^B(\tau)^2=s^2$. The transformation (4) should be thought of as
a change of variables in the path integral (5) with Jacobian equal to one.
After this transformation, the spin vectors $\vec S_i^A$ and $\vec S_i^B$
depend on the complex fields $\alpha_i^+(\tau),\alpha_i(\tau)$ and
$\beta_i^+(\tau),\beta_i(\tau)$.

I shall define the spin-vectors $\vec S_{(1)i}$ by the equations
\FL
\begin{equation}
\left. \vec S^A_i \right|_{\beta_i^+=0,\beta_i=0}=
\left. \vec S^B_i \right|_{\beta_i^+=0,\beta_i=0}=
\frac {1} {2} \vec S_{(1)i}(\alpha^+,\alpha)
\end{equation}
They depend on the slow modes and satisfy $\vec S_{(1)i}^2=(2s)^2$.

It is convenient to represent the spin-vectors in the form
\FL
\begin{equation}
\vec S_i^A =\frac 12\sqrt {1-\frac {\bar\varphi_i^A\varphi_i^A}{s^2}}
\vec S_{(1)i}+\bar\varphi_i^A\vec {\text{\large e}}_i+
\vec {\bar {\text {\large e}}}_i\varphi_i^A\,\,; \qquad
\vec S_i^B = \frac 12\sqrt {1-\frac {\bar\varphi_i^B\varphi_i^B}{s^2}}
\vec S_{(1)i}+\bar\varphi_i^B\vec {\text{\large e}}_i+
\vec {\bar {\text {\large e}}}_i\varphi_i^B
\end{equation}
where the real vectors
$\vec {\text{\large e}}_j^1=\vec {\text{\large e}}_j+
\vec {\bar {\text {\large e}}}_j$ and
$\vec {\text{\large e}}_j^2=\frac 1i\left(\vec {\text{\large e}}_j-
\vec {\bar {\text {\large e}}}_j\right)$
depend on the fields $\alpha_j^+(\tau)$ and $\alpha_j(\tau)$. They are
orthogonal to the vectors $\vec S_{(1)j}$ and to each other and
$\vec {\bar {\text {\large e}}}_j\cdot\vec {\text{\large e}}_j=\frac 12$

The representation (7) is not unique. There is an arbitrariness in
the definition of the vectors $\vec {\bar {\text {\large e}}}_j$,
$\vec {\text{\large e}}_j$ and the coefficients
$\bar\varphi_j^A,\varphi_j^A$,
$\bar\varphi_j^B,\varphi_j^B$. It is clear from Eqs (7) that a
local $U(1)$ transformation of the vectors and the coefficients
does not change the spin vectors $\vec S_i^A$ and $\vec S_i^B$. I shall use
the following representation for the vectors $\vec {\text{\large e}}_j$ and
$\vec {\bar {\text {\large e}}}_j$
\FL
\begin{eqnarray}
\text{\large e}_i^+ & = & -\frac 1{4s}\alpha_i\alpha_i \hspace{0.5cm}
\text {\large e}_i^- = \frac 1{4s}(4s-\alpha_i^+\alpha_i) \hspace{0.5cm}
\text{\large e}_j^3 = -\frac 1{4s}(4s-\alpha_i^+\alpha_i)^{\frac 12}\alpha_i
\nonumber \\
\bar {\text {\large e}}_i^+ & = & \frac 1{4s}(4s-\alpha_i^+\alpha_i)
\hspace{0.5cm}
\bar {\text{\large e}}_i^- = -\frac 1{4s}\alpha_i^+\alpha_i^+ \hspace{0.5cm}
\bar {\text{\large e}}_j^3 =
-\frac 1{4s}\alpha_i^+(4s-\alpha_i^+\alpha_i)^{\frac 12}
\end{eqnarray}

The next step of the renormalization group approach is the elimination of
the fast modes. One can integrate over the fields $\beta_i^+(\tau)$ and
$\beta_i(\tau)$ only perturbatively. For this reason I expand $\varphi_i^A$
and $\varphi_i^B$ in powers of $\beta_i^+(\tau)$ and $\beta_i(\tau)$ fields.
Calculating the coefficients in the expansion to leading order of $s^{-1}$,
one obtains
\FL
\begin{equation}
\varphi_i^A \simeq \sqrt {s}\beta_i+\frac {\alpha_i}{4\sqrt {s}}
\beta_i^+\beta_i  \hspace{0.5cm}
\varphi_i^B \simeq -\sqrt {s}\beta_i+\frac {\alpha_i}{4\sqrt {s}}
\beta_i^+\beta_i
\end{equation}
where the terms proportional to $\beta_i\beta_i$ are dropped because they
do not contribute to the effective action. Substituting (9) and (7) in
the hamiltonian (2) and selecting the quadratic terms in $\beta_i^+$ and
$\beta_i$ one gets
\FL
\begin{eqnarray}
h & = & -\frac J2\sum\limits_{\nu}\sum\limits_{i\in L_1}\vec S_{(1)i}\cdot
\vec S_{(1)j} +
sJ\sum\limits_{\mu}\sum\limits_{i\in L_1}\left (\beta_i^+\beta_i+
\beta_j^+\beta_j+ \beta_i^+\beta_j+ \beta_j^+\beta_i\right) \nonumber \\
& & -\frac {J}{4\sqrt {s}}\sum\limits_{\nu}\sum\limits_{i\in L_1}\left [
\vec S_{(1)j}\cdot \left (\vec {\text{\large e}}_i\alpha_i^++
\vec {\bar {\text {\large e}}}_i\alpha_i\right)\beta_i^+\beta_i+
\vec S_{(1)i}\cdot \left (\vec {\text{\large e}}_j\alpha_j^++
\vec {\bar {\text {\large e}}}_j\alpha_j\right)\beta_j^+\beta_j\right]
\nonumber \\
& & -\frac {J}{8s}\sum\limits_{\nu}\sum\limits_{i\in L_1}
\left (\vec S_{(1)i}-\vec S_{(1)j}\right)^2\left (\beta_i^+\beta_i
+\beta_j^+\beta_j\right)  \\
& & -sJ\sum\limits_{\nu}\sum\limits_{i\in L_1}
\left [\left (1-2\vec {\bar {\text {\large e}}}_j\vec {\text{\large e}}_i
\right)\beta_i^+\beta_j+
\left (1-2\vec {\bar {\text {\large e}}}_i\vec {\text{\large e}}_j
\right)\beta_j^+\beta_i\right] \nonumber
\end{eqnarray}
where $j=i+a_{\nu}$ and the terms proportional to $\beta_i\beta_j$ and
$\beta_i^+\beta_j^+$ are dropped again.

Integrating over the fields $\beta_i^+(\tau)$ and $\beta_i(\tau)$ we lose
the rotational invariance of the theory. This is well known disadvantage
of the spin-wave approximation. To restore the rotational symmetry, and to
obtain a rotationally invariant effective action one has to average it over
the elements of the group of rotation \cite{3}. Under the nonlinear
rotational transformations of the fields $\alpha_i^+$ and $\alpha_i$
the scalar product of the vectors $\vec {\bar {\text {\large e}}}_i$ and
$\vec {\text{\large e}}_j$ changes the phase
$\vec {\bar {\text {\large e}}}_i\cdot\vec {\text{\large e}}_j
\to \exp[i(\delta_i-\delta_j)]
\vec {\bar {\text {\large e}}}_i\cdot\vec {\text{\large e}}_j$,
where $\delta_i$ depends on the fields $\alpha_i^+$ and $\alpha_i$
and on the group elements.
Then the module of the scalar product is invariant and the phase
$\gamma_{ij}$ transforms as an $U(1)$ gauge field on a square lattice
$\gamma_{ij}^{\prime}=\gamma_{ij}+\delta_i-\delta_j$.
Hence, the phases enter the invariant action in the same way
as $U(1)$ gauge fields enter a gauge invariant action. It is not difficult
to check that these terms have higher dimension and one has to drop them.
This means to replace
$\vec {\bar {\text {\large e}}}_i\cdot\vec {\text{\large e}}_j$ and
$\vec {\bar {\text {\large e}}}_j\cdot\vec {\text{\large e}}_i$ by
$|\vec {\bar {\text {\large e}}}_i\cdot\vec {\text{\large e}}_j|$
in the hamiltonian (10).
Straightforward calculation using (8) leads to the result
\FL
\begin{equation}
|\vec {\bar {\text {\large e}}}_i\cdot\vec {\text{\large e}}_j|=
\frac 12-\frac {1}{32s^2}\left (\vec S_{(1)i}-\vec S_{(1)j}\right)^2
\end{equation}
I shall treat  the other terms in the same way.
Using the representation (8) and (6), one gets
\FL
\begin{equation}
\vec S_{(1)j}\cdot \left (\vec {\text{\large e}}_i\alpha_i^++
\vec {\bar {\text {\large e}}}_i\alpha_i\right)+
\vec S_{(1)i}\cdot \left (\vec {\text{\large e}}_j\alpha_j^++
\vec {\bar {\text {\large e}}}_j\alpha_j\right)=-\frac {1}{2\sqrt {s}}
\left (\vec S_{(1)i}-\vec S_{(1)j}\right)^2+\dots
\end{equation}
where the dots stand for terms which become irrelevant after group averaging.

Integrating over the fast modes $\beta_i^+(\tau)$ and $\beta_i(\tau)$
only the tadpole graphs contribute to the one-loop approximation.
Calculating them at zero temperature and taking into account the Eqs (11)
and (12) I obtain the effective hamiltonian
$h_{\text {eff}}=-J^{\prime}\sum\limits_{\nu}\sum\limits_{i\in L_1}
\vec S_{(1)i}\cdot\vec S_{(1)j}$,
where $J^{\prime}=\frac 12J\left[1-\frac {1}{2s}\left(\frac 12+\frac {1}{\pi^2}
\right)\right]$ and $j=i+a_{\nu}$. The sum is over the sites of the
lattice $L_1$ and over the two directions of the previous lattice. I want to
represent the sum as a sum over the nearest neighbours of the lattice $L_1$.
For this reason I write the hamiltonian in momentum representation
$h_{\text {eff}}=-J^{\prime}{\sum\limits_q}^{\prime}
\vec S_{(1)}{}_{-q}\cdot \vec S_{(1)}{}_q
\left(\cos aq_x+\cos aq_y\right)$.
Let us rotate the reduced Brillouin zone
$q_x=\frac {1}{\sqrt 2}\left(k_x+k_y\right), \
q_y=\frac {1}{\sqrt 2}\left(k_x-k_y\right)$
The new wave-vectors $k_x$ and $k_y$ run over the Brillouin zone of the lattice
$L_1$. One can obtain the relation
$\cos aq_x+\cos aq_y=\frac12 \left(2+\cos \sqrt 2ak_x+\cos \sqrt 2ak_y
\right)+0(k^4)$.
Substituting it, and rewriting the expression for the effective
hamiltonian in site representation one obtains
$h_{(1)}=-J_1\sum\limits_{<i,j>}\vec S_{(1)i}\cdot\vec S_{(1)j}$.

The effective hamiltonian $h_{(1)}$ is a hamiltonian of a Heisenberg ferromagnet
with spin $s_1=2s$ (see eq (6)), defined on a square lattice with spacing
$a_1=\sqrt 2 a$, and with exchange constant
$J_1=\frac14 J\left[1-\frac {1}{2s}\left(\frac 12+
\frac {2}{\pi^2}\right)\right]$.

Repeating the procedure one obtains the recursion equation
\FL
\begin{equation}
s_{n+1} = 2s_n, \hspace{0.5cm}
a_{n+1} = \sqrt 2 a_n, \hspace{0.5cm}
J_{n+1} = \frac 14 J_n\left[1-\frac {1}{2s_n}\left(\frac 12+\frac {2}{\pi^2}
\right)\right]
\end{equation}

It is readily to solve them
\FL
\begin{equation}
s_n = 2^ns, \hspace{0.5cm}
a_n = 2^{\frac n2} a, \hspace{0.5cm}
J_n = \frac {1}{4^n} J\prod\limits_{l=1}^n
\left[1-\frac {1}{2^{l}\,s}\left(\frac 12+\frac {2}{\pi^2}
\right)\right]=\frac {1}{4^n} J \rho_n(s)
\end{equation}

I shall consider the equation for the chemical potential. Making the
transformation (4) and rewriting it in momentum space representation one
obtains
\FL
\begin{equation}
\frac {1}{N_1}\sum\limits_q\left[\langle \alpha_q^+\alpha_q \rangle +
\langle \beta_q^+\beta_q \rangle \right]=2s=s_1
\end{equation}
where $\vec q$ runs the Brillouin zone of $L_1$ lattice. The dispersion of
$\beta_q$ bosons is positive, hence, when the temperature goes to zero,
$\langle \beta_q^+\beta_q \rangle$ goes to zero exponentialy and can be
dropped
in comparisson to $\langle \alpha_q^+\alpha_q \rangle$. As a result the
equation for the chemical potential in low-temperature limit takes the form
$\frac {1}{N_1}\sum\limits_q \langle \alpha_q^+\alpha_q \rangle =s_1$.
In the limit of infinitely many recursions, one can calculate
$<\alpha_q^+\alpha_q>$ in the spin-wave approximation [3]. Then
\FL
\begin{equation}
\mu(\beta\to\infty)=\frac {1}{\beta}e^{-4\pi s^2J\rho(s)\beta}
\end{equation}
where
\FL
\begin{equation}
\rho(s)=\prod\limits_{l=1}^{\infty}\left[1-\frac {1}{2^l\,s}\left(
\frac 12+\frac {2}{\pi^2}\right)\right]
\end{equation}

To obtain the correlation length one has to consider the asymptotic
behaviour of the two-point static correlation function at long distance.
In momentum space this translates to small wave-vectors. Each recursion
separates smaller and smaller area around the origin of the Brillouin zone.
In the limit of infinitely many recursions the correlation function can
be calculated in the spin-wave approximation. Using the relation between the
chemical potential and the correlation length in spin-wave theory [3],
I obtain the low-temperature asymptotic of the correlation length
\FL
\begin{equation}
\xi(\beta\to\infty)=\lim_{n\to\infty}\frac 12
\left(\frac {s_nJ_na_n^2}{\mu(\beta)}\right)^{\frac 12}=
\frac 12a\sqrt {sJ\rho(s)\beta}e^{2\pi s^2J\rho(s)\beta}
\end{equation}

Let us add a new (Zeeman) term in the hamiltonian
$h\to h-2H\sum\limits_i\hat S_i^3$,
where $H$ is an external constant magnetic field. Then the uniform
susceptibility is proportional to
$\left. d^2{\cal F}/dH^2 \right|_{H=0}$,
where ${\cal F}$ is the free energy of the system. The recursion does not
change the magnetic field. For the free energy one obtains
${\cal F} = \frac {1}{2} {\cal F}_{1} + \dots $,
where ${\cal F}_{1}$ is the free energy of the system which is defined by the
hamiltonian $h_{(1)}$  with an additional Zeeman term. The dots stand for
terms independent on the magnetic field. The multiplier $\frac 12$ in front
of ${\cal F}_{1}$ is a consequence of the fact that the number of sites of
the lattice $L_1$ is $N_1=\frac {1}{2} N$. Hence
$\chi(\beta\to\infty)=\frac {1}{2} \chi_1(\beta\to\infty)$
and to leading order of small temperature
\FL
\begin{equation}
\chi(\beta\to\infty)=\lim \limits_{n\to\infty} \frac
{1}{2^n}\chi_n(\beta\to\infty)=
\frac {1}{3\pi sJ\rho(s)}e^{4\pi s^2J\rho(s)\beta}
\end{equation}
where $\chi_n(\beta\to\infty)$ is calculated in the spin-wave approximation [3].

From Eq.(17) follows that $\rho(s\to\infty)=1$. Substituting this result
in formulas (16),(18) and (19), one gets the results from the
spin-wave theory [3]. On the other hand
$\rho(s)_{|s=\frac 12}=0.1327 \qquad \rho(s)_{|s=1}=0.4462$
and one can see that the results for small spins are essentialy
different from those in the spin-wave theory.

In conclusion I shall summarize what has been discussed here. Ideas and tools
developed in the papers of K.G.Wilson \cite{5}, A.A.Migdal \cite{6},
L.P.Kadanoff \cite{7}, A.M.Polyakov \cite{8}, D.Nelson, R.A.Pelcovits,
S.Chakravarty and B.I.Halperin \cite{9,10} have been adapted
to formulate a renormalization group approach in the
spin-wave theory. The formalism allows the analysis of systems with
small spin.

\acknowledgements

This work was supported by the Bulgarian National Science Foundation
under Grant No 92-F214.

\end{document}